
%
%
%
%
\input amstex

\define\a{\alpha}

\redefine\b{\beta}

\redefine\o{\omega}

\redefine\l{\lambda}

\redefine\top{\text{top}}
\define\index{\text{index}}
\define\RM{\Bbb R}

\define\gm{\bold g}

\define\<#1,#2>{\langle #1,#2\rangle}
\define\TR{\text{tr}}
\define\dep(#1,#2){\text{det}_{#1}#2}
\define\norm(#1,#2){\parallel #1\parallel_{#2}}
\magnification\magstep1
\documentstyle{amsppt}
\topmatter
\title INDEX THEORY, GERBES, AND HAMILTONIAN QUANTIZATION \endtitle
\author Alan Carey$^1$, Jouko Mickelsson$^2$, and Michael Murray$^1$
\endauthor
\affil  Department of Mathematics, University of Adelaide, Adelaide,
SA 5005,
Australia$^1$;
Department of Theoretical Physics, Royal Institute of Technology,
S-10044 Stockholm, Sweden. e-mail: jouko\@theophys.kth.se $^2$ \endaffil

\endtopmatter

\document
\noindent{ABSTRACT.}  We give an Atiyah-Patodi-Singer index theory
construction of the bundle of fermionic Fock spaces parametrized by vector
potentials in odd space dimensions and prove that this leads in a simple
manner to the known Schwinger terms (Faddeev-Mickelsson cocycle) for the
gauge group action. We relate the APS construction to the bundle gerbe
approach discussed recently by Carey and Murray, including an explicit
computation of the Dixmier-Douady class. An advantage of our method is
that it can be applied whenever one has a form of the APS theorem at hand,
as in the case of fermions in an external gravitational field.

\vskip 0.5 in
\noindent{1. INTRODUCTION.}
\vskip 0.3in

There are subtleties in defining the fermionic Fock spaces in the case of
chiral (Weyl) fermions in external vector potentials. The difficulty is
related to the fact that the splitting of the one particle fermionic
Hilbert space $H$ into positive and negative energies is not continuous as
a function of the external field. One can easily construct paths in the
space of external fields such that at some point on the path a positive
energy  state dives into the negative energy space (or vice versa). These
points are obviously discontinuities in the definition of the space of
negative energy states and therefore the fermionic vacua do not form a
smooth vector bundle over the space of external fields. This problem does
not arise if we have massive fermions in the temporal gauge $A_0=0.$ In
that case there is a mass gap $[-m,m]$ in the spectrum of the Dirac
hamiltonians and the polarization to positive and negative energy
subspaces is indeed continuous.

If $\l$ is a real number not in the spectrum of the hamiltonian then one
can define a bundle of fermionic Fock spaces $\Cal F_{A,\l}$ over the set
$U_{\l}$ of external fields $A,$ $\l\notin Spec(D_A).$ It turns out that
the Fock spaces $\Cal F_{A,\l}$ and $\Cal F_{A,\l'}$ are naturally
isomorphic up to a phase. The phase is related to the arbitrariness in
filling the Dirac sea between vacuum levels $\l,\l'.$ In order to
compensate this ambiguity one defines a tensor product $\Cal
F'_{A,\l}=\Cal F_{A,\l} \otimes DET_{A,\l},$ where the second factor is a
complex line bundle over $U_{\l}.$ By a suitable choice of the determinant
bundle the tensor product becomes independent of $\l$ and one has a
well-defined bundle $\Cal F'$ of Fock spaces over all of $\Cal A.$

Next one can ask what is the action of the gauge group on $\Cal F'.$ The
gauge action in $U_{\l}$ lifts naturally to $\Cal F.$ Thus the only
problem is to construct a lift of the action on the base to the total
space of $DET_{\l}.$ Note that the determinant bundle here is a bundle
over external fields in \it odd dimension, \rm and therefore one would
expect that it is trivial (curvature equal to zero) on the basis of the
families index theorem. However, it turns out that the relevant
determinant bundle actually comes from a determinant bundle in even
dimensions. Instead of single vector potentials we must study paths in
$\Cal A,$ thus the extra dimension. The relevant index theorem is then the
Atiyah Patodi Singer (APS) index theorem for even dimensional manifolds
with a boundary [AtPaSi]; physically, the boundary can be interpreted as
the union of the space at the present time and in the infinite past.

The gauge action in the bundle $\Cal F$ leads to Schwinger terms in the
Lie algebra commutation relations of the gauge currents. These commutator
anomalies have been discussed before in the literature from different
points of view. In this paper we give a simple derivation using the
families index theorem, giving a Fock space formulation for the descent
equations leading from the space-time anomalies to hamiltonian anomalies.
We also explain the relation between the Schwinger terms and the
Dixmier-Douady class (which is a certain closed 3-form on the moduli space
of gauge connections) in de Rham cohomology.

\vskip 0.3in
\noindent{2. THE ODD DETERMINANT BUNDLES.}
\vskip 0.3in

Let $M$ be a smooth compact manifold without boundary equipped with a
spin structure. We assume that the dimension of $M$ is odd and equal to
$2n+1.$ Let $S$ be the spin bundle over $M,$ with fiber isomorphic
to  $\Bbb C^{2^n}.$ Let $H$ be the space of square integrable sections
of the complex vector bundle $S\otimes V,$ where $V$ is a trivial vector
bundle over $M$ with fiber to be denoted by the same symbol $V.$ The measure
is defined by a fixed metric on $M$ and $V.$ We assume
that a unitary representation $\rho$ of a compact group $G$ is given in the
fiber. The set of smooth vector potentials on $M$ with values in the Lie
algebra $\gm$ of $G$ is denoted by $\Cal A$ or $\Cal A_{2n+1},$ depending
on whether there is a chance of confusion.

For each $A\in \Cal A$ there is a massless hermitean Dirac operator $D_A.$
Fix a
potential $A_0$ such that $D_A$ does not have the zero as an eigenvalue
and let $H_+$ be the closed
subspace spanned by eigenvectors belonging to positive
eigenvalues of $D_{A_0}$ and $H_-$ its orthogonal complement.
More generally for any potential $A$ and any real $\l$ not belonging
to the spectrum of
$D_A$ we define the spectral decomposition $H=H_+(A,\l)\oplus H_-(A,\l)$
with respect to the operator $D_A - \l.$ Let ${\Cal A}_0$ denote the set of
all pairs $(A,\l)$ as above and let $U_{\l}=\{A\in\Cal A| (A,\l)\in{\Cal
A}_0\}.$

Over the set $U_{\l\l'}=U_{\l}\cap U_{\l'}$ there is a canonical complex
line bundle, to be denoted by $DET_{\l\l'}.$ Its fiber at $A\in U_{\l\l'}$
is the top exterior power
$$
DET_{\l\l'}(A) = \wedge^{\top} (H_+(A,\l)\cap H_-(A,\l'))\tag2.1
$$
where we have assumed $\l<\l'.$ For completeness we put $DET_{\l\l'}=
DET_{\l'\l}^{-1}.$ Since $M$ is compact, the spectral subspace corresponding
to the interval $[\l,\l']$ in the spectrum is finite-dimensional and the
complex line above is well-defined.

It is known [Mi1, CaMu1] that there exists a complex line bundle $DET_{\l}$
over each of the sets $U_{\l}$ such that
$$
DET_{\l'}= DET_{\l}\otimes DET_{\l\l'} \tag2.2
$$
over the set $U_{\l\l'}.$  In [CaMu, CaMu1] the structure of these line
bundles was studied with the help of bundle gerbes. In particular, there
is an obstruction for passing to the quotient by the group $\Cal G$ of
gauge transformations which is  given by the Dixmier-Douady class of the
gerbe. (In [Mi1] the structure of the bundles and their relation to
anomalies was found by using certain embeddings to infinite-dimensional
Grassmannians.)

In this paper we shall compute the curvature of the (odd dimensional)
determinant bundles from Atiyah-Patodi-Singer index theory and  we obtain
the Schwinger terms in the Fock bundle directly from the local part of the
index density.

To each $(A,\l)$ in ${\Cal A}_0$ we associate a euclidean Dirac operator
on the $2n+2$ dimensional manifold $M\times [0,1]$ with the obvious metric
and spin structure. This Dirac operator is $$D^{(2n+2)}_{A(t)}=
\frac{\partial}{\partial t} +D_{A(t)}\tag2.3$$ where the time dependent
potential is $A(t)= f(t)A+(1-f(t))A_0.$ Here $f$ is a  fixed smooth real
valued function on the interval $[0,1]$ such that $f(0)=0, f(1) =1,$  and
the function is constant near the end points. It turns out that the choice
of $f$ does not influence our results as we show at the end of the
section.

We fix the boundary conditions for $D^{(2n+2)}_{A(t)}$ such that at the
boundary component $t=0$ the spinor fields should belong to $H_-$ whereas
at $t=1$ the spinor field is in $H_+(A,\l).$ This type of boundary
condition was used in [AtPaSi] in the proof of index theorems (in even
dimensions) when the manifold has a boundary. The Dirac operator is
nonhermitean, it is really a map between two different spaces, namely the
space of left handed spinors $S_-$ and right handed spinors $S_+.$ The
kernel and cokernel of $D^{(2n+2)}_{A(t)}$ are finite dimensional vector
spaces.

The tensor product of the top exterior powers of the dual of the kernel
and the cokernel of $D^{(2n+2)}_{A(t)}$ defines a complex line
$DET_{\l}(A).$ Together these lines define a complex line bundle
$DET_{\l}$ over $U_{\l},$ the set of potentials not having $\l$ as an
eigenvalue. The bundle does not extend to all of $\Cal A$ since the
boundary conditions change abruptly at points in the parameter space such
that the corresponding boundary Dirac operator has zero modes.

There is an important alternative description of the determinant line
bundle. Let $\{\psi_n\}$ be a basis of eigenvectors at the boundary
component $t=1$ corresponding to eigenvalues $\l_n >\l,$
$$
D_A \psi_n = \l_n \psi_n.
$$
The nonhermitean time evolution
$$
i\partial_t \phi = -i D_{A(t)} \phi\tag2.4
$$
defines for each $n$ a unique solution $\phi_n$ on $M\times [0,1]$ such
that at $t=1$ $\phi_n(x,1)=\psi_n(x).$ The vectors $\phi_n(x,0)$ span an
infinite dimensional plane $W=W(A,\l)$ in $H.$ Let $\pi_+$ be the
projection from $W$ to $H_+.$ The kernel of this projection can be
identified as the kernel of $D^{(2n+2)}_{A(t)}$ through restriction to the
boundary $t=0.$ Similarly, the cokernel of $D^{(2n+2)}_{A(t)}$ is
identified as the cokernel of $\pi_+.$ This is because the boundary
conditions for the adjoint operator $(\partial_t +D_{A(t)})^*= -\partial_t
+D_{A(t)}$ are orthogonal to the boundary conditions of
$D^{(2n+2)}_{A(t)},$ [AtPaSi]. At $t=0$ the vectors in the domain of the
adjoint belong to $H_+$ whereas at $t=1$ they belong to $H_-(A,\l).$ On
the other hand, coker$\, \pi_+= W^{\perp}\cap H_+$ and a zero mode of the
adjoint is orthogonal to a zero mode of $D^{(2n+2)}_{A(t)}$ at $t=0.$ Thus
$$
DET_{\l}(A) = \wedge^{\top}(\text{ker} \pi_+)^* \otimes
\wedge^{top}(\text{coker}
\pi_+).\tag2.5
$$

We choose an orthonormal basis $\{e_n\}_{n\in\Bbb Z}$ such that the vectors
with a nonnegative $n$ belong to $H_+$ and those with a negative $n$
belong to $H_-.$  Since $\pi_+$ is a Fredholm operator, of index $k=
\text{dim ker} \pi_+- \text{dim coker} \pi_+$ say, the projection
$\pi_{+,k}$ from $W$ to the plane $H_k$ spanned by the vectors $\{e_n\}_
{n\geq-k}$ is almost invertible, i.e., there is a linear map $q: H_k\to W$
such that $q\pi_+$ and $\pi_+ q$ differ from the identity operator by a
finite rank operator. The pseudo-inverse $q$ is fixed by a choice of basis
$\{u_1,\dots, u_r\}$ in  ker$\pi_+$ and a basis $\{v_1,\dots,v_{r-k}\}$ in
coker$\pi_+:$ The map $\pi_+$ gives an isomorphism between
$\pi_+(W)\subset H_+$ and (ker$\pi_+)^{\perp}\cap W.$ This isomorphism is
complemented to an isomorphism between $W$ and $H_k$ by adjoining to
$u_i$ the vector $v_i$ for $i=1,2,\dots ,r-k$ and $u_i\mapsto e_{i-r}$ for
$i=r-k+1,\dots,r,$ when $k$ is nonnegative. When $k<0$ we define $H_k$ as
the space spanned by $e_i$ with $i \geq -k$ and proceed as before.

The image $\{w_{-k},w_{-k+1},\dots\}$ of the basis of $H_k$ under $q$ is
\it an admissible basis \rm of $W,$ [PrSe]. By definition, any admissible
bases of $W$ is a basis obtained from $\{w_i\}$ by a unitary rotation by
an operator $1+R,$ where $R$ is trace-class. The operators $1+R$ have a
well-defined determinant. Over $W(A,\l)$ (that is, over $A\in U_{\l}$)
there is a complex line defined as the set of all admissible basis of $W$
modulo basis transformations by operators with unit determinant. As we saw
above, the ambiguity in the construction of an admissible basis is the
same as the freedom of choosing the basis in ker$\pi_+$ and coker$\pi_+.$
It follows that the determinant line is naturally identified as the
complex line in the Pressley-Segal construction.

Any choice $\{f_n\}$ of a basis of eigenvectors of $D_A$ corresponding to
eigenvalues in the interval $[\l,\mu]$ gives now an isomorphism between
the determinant lines $DET_{\l}(A)$ and $DET_{\mu}(A).$ Namely, an
admissible basis $\{w_n\}$ of $W(A,\mu)$ can be completed to an admissible
basis of $W(A,\l)$ by adding the time evolved vectors obtained from
$\{f_n\}$ by the euclidean time evolution backwards in time from $t=1$ to
$t=0.$ Clearly, a rotation $R $ of the basis $\{f_n\}$ induces a rotation
of the determinant line $DET_{\l}(A)$ by a phase equal to det$R.$ On the
other hand, a choice of the basis $\{f_n\}$ modulo unitary transformations
$R$ with det$R=1$ is equivalent to choosing an element in the complex line
$DET_{\l\mu}(A).$ This shows that we can identify
$$DET_{\mu}(A)= DET_{\l\mu}(A) \otimes DET_{\l}(A),$$
as already stated in (2.2).

An alternative proof of this result can be given which uses the APS index
theorem as follows.  Denote $W_+(A,\l)=W(A,\l)$ and
$W_-(A,\l)=W(A,\l)^{\perp}.$ Define
$$
K(A, \lambda) =   W_+(A, \lambda) \cap H_-   \quad \text{and}\quad
K(A, \lambda') =   W_+(A, \lambda') \cap H_-
$$
and
$$
C(A, \l) = H_+ \cap   W_-(A, \lambda) \quad \text{and}\quad
C(A, \l') = H_+ \cap   W_-(A, \lambda').
$$

These are the kernels and co-kernels of the even dimensional
Dirac operators formed out of $A_0$ and $A$ with projection
at $t=1$ onto the eigenspaces greater than $\lambda'$ and $\lambda$
respectively. So we have
$$
DET_\lambda = \wedge^{\top}(K(A, \lambda)^* \oplus C(A, \lambda))
\quad\hbox{and}\quad
DET_{\lambda'} = \wedge^{\top}(K(A, \lambda')^* \oplus C(A, \lambda') ).
$$

Recall that
$$
H = W_-(A, \l) \oplus (W_+(A,\l)\cap W_-(A,\l')) \oplus W_+(A, \l')
$$
so that we have
$$
W_-(A, \l') = W_-(A, \l) \oplus W_+(A,\l)\cap W_-(A,\l')
$$
and
$$
W_+(A, \l) = W_+(A, \l') \oplus W_+(A,\l)\cap W_-(A,\l').
$$
So orthogonal projection defines a map
$$
K(A, \lambda)/ K(A, \lambda') \to W_+(A,\l)\cap W_-(A,\l')
$$
and similarly
$$
C(A, \l') / C(A, \l) \to W_+(A,\l)\cap W_-(A,\l').
$$
Adding  these  gives a map
$$
K(A, \lambda)/ K(A, \lambda') \oplus C(A, \l') / C(A, \l)  \to
W_+(A,\l)\cap W_-(A,\l')
$$
If we can prove that this final map is an isomorphism then by
wedging to the top power on either side we will have constructed
an isomorphism
$$
DET_{\lambda'}(A)\otimes   DET_\lambda(A)^* = DET_{\lambda\l'}(A).
$$
which gives the desired result in equation (2.2).
It is easy to prove that this map is injective because the
images of the two factors are, in fact, orthogonal. It remains
to do surjectivity and this comes from a dimension count
which follows from the APS index theorem.
It suffices to show that
$$
\dim (K_\l) - \dim( K_{\lambda'}) +  \dim( C_{\lambda'}) - \dim( C_\l)
= \dim(W_+(A,\l)\cap W_-(A,\l')).
$$
Given $(A_0, 0)$ and $(A, \lambda) $ let $D[(A_0, 0), (A, \lambda)]$ be
the four dimensional Dirac operator as above. We need to prove then that
$$
\index(D[(A_0, 0), (A, \lambda')] - \index(D[(A_0, 0), (A, \l)] =  \dim(
W_+(A,\l')\cap W_-(A,\l)).
$$
It is easy to show that
$$
\dim( W_+(A,\l)\cap W_-(A,\l'))= \dim( H_+(A,\l)\cap H_-(A,\l'))
= \index{(D[(A, \lambda'), (A, \l)]}
$$
so the result follows from the fact that the index is additive. That
is
$$
\index(D[(A, \l), (B, \l')] +  \index(D[(B , \l'), (C, \l'')] =
\index(D[(A , \l), (C, \l'')]).
$$
This additivity of the index is a direct consequence of the index theorem
itself. The index is a sum of two terms. The first is an integral of a \it
local \rm differential polynomial of the vector potential and therefore it
is manifestly additive in time. The second term is also additive because
it is equal to $\frac12(\eta(t=1)-\eta(t=0)).$ On the common boundary the
eta invariants (for the boundary operator $B$) in index$(D[(A,\l),(B,
\lambda')] +$ index$(D[(B, \lambda'),(C,\l'')]$ cancel.

The real parameters $\l,\lambda',\l''$ do not change the discussion since
we can always consider  operators such as
$$
D_{A(t)} - f(t)\l -(1-f(t))\l'
$$
instead of $D_{A(t)}.$  For the shifted Dirac operators we can use the \lq
vacuum level' value 0.

Finally, the geometry of the determinant bundles $DET_{\l}$ is described
by the families index theorem. Normally, the determinant bundle over $\Cal
A$ in even dimensions is trivial whereas the bundle over the moduli space
of gauge orbits $\Cal A/\Cal G_0$ is nontrivial. Here $\Cal G_0$ is the
group of based gauge transformations, $g(p)= 1,$ where $p\in M$ is some
fixed point. However, in the present case we are studying potentials on a
manifold with boundary and the boundary conditions depend globally on the
potential $A$ not having $\l$ as a zero mode. The parameter space is not
affine, the determinant bundle is nontrivial. (We stress again that the
determinant bundle $DET_{\l}$ does \it not \rm extend to the space of all
vector potentials; there are discontinuities at the points $A$ for which
$\l$ is an eigenvalue.)

We use the same APS boundary conditions for the operators
$D^{(2n+2)}_{A(t)}$ as before. Then according to [AtPaSi],
$$
\text{index} D^{(2n+2)}_{A(t)}= \int Ch(A(t)) -\frac12 (\eta(t=1) -
\eta(t=0)) \tag2.6
$$
assuming that the boundary operators do not have zero modes. Here $Ch$ is a
characteristic class depending in general on the vector potential and the
metric.  This term is the same as in the case of a manifold without a
boundary. The eta invariant $\eta(D_A)$ for a hermitean operator is
defined through analytic continuation of $$  \eta_s(A)= \sum_i
\frac{\l_i}{|\l_i|^s}$$ which is well-defined for $s>>0,$ to the point
$s=1,$ where the $\l_i$'s are the eigenvalues of $D_A.$ The $\eta$
-invariant term in (2.6) depends only on data on the boundary.

The Chern class of the determinant bundle $DET$ over this class of Dirac
operators in completely determined by integrating the corresponding de
Rham form over two dimensional cycles $S^2 \mapsto$ set of Dirac
operators.

We recall some facts about lifting a group action on the base space $X$ of
a complex line bundle to the total space $E.$ Let $\omega$ be the
curvature 2-form of the line bundle. It is integral in the sense that
$\int \omega$ over any cycle is $2\pi \times$ an integer. Let $G$ be a
group acting smoothly on $X.$ Then there is an extension $\hat G$ which
acts on $E$ and covers the $G$ action on $X.$ The fiber of $\hat G \to G$
is equal to $Map(X,S^1).$ As a vector space, the Lie algebra of the
extension is $\gm \oplus Map(X,i\Bbb R).$ The commutators are defined as
$$[(a,\a),(b,\b)]= ([a,b],\o(a,b)+\Cal L_a \b-\Cal L_b \a)\tag2.7$$ where
$a,b\in \gm$ and $\a,\b: X\to i\Bbb R.$ The vector fields generated by the
$G$ action on $X$ are denoted by the same symbols as the Lie algebra
elements $a,b;$ thus $\o(a,b)$ is the function on $X$ obtained by
evaluating the 2-form $\o$ along the vector fields $a,b.$ The Jacobi
identity $$\o([a,b],c) + \Cal L_a \o(b,c) + \text{cyclic permutations}
=0$$ for the Lie algebra extension $\hat\gm$ follows from $d\o=0.$

For the computation of the Schwinger term we need only the curvature along
gauge directions for the boundary operator $D^{(2n+1)}_{A(t=1)}.$
According to the general theory of determinant bundles: the integral of
the first Chern class over a $S^2$ in the parameter space of Dirac
operators = the index of the family of Dirac operators. That means: one
has to choose (any) connection on $B=S^2\times [0,1]\times M$ such that
along $[0,1]\times M$ it is equal to the potential
$f(t)A(x,z)+(1-f(t))A_0$ (here $z\in S^2$ parametrizes the family of
operators) and satisfies the appropriate boundary conditions. The
appropriate Dirac operator is then the operator $D_B$ on $B$ related to
this connection.

Consider a family of gauge transformed potentials $A(x,z)= g A g^{-1} +
d_x g g^{-1},$ where $x\mapsto g(x,z)$ is a family of gauge
transformations parametrized by points $z\in S^2.$ To this family of
potentials we associate a Dirac operator $D_B$ on $B.$ Formally $$D_B =
D(S^2) + D^{(2n+2)}_{A(t)} + f(t) \rho(z)^{-1}\gamma^z \cdot \partial_z
\rho(z)\tag2.8$$ where the first term is the Dirac operator on $S^2$
determined by a metric and fixed spin structure; $\gamma^z$ stand for a
pair of gamma matrices to the $S^2$ directions. The boundary conditions at
$t=1$ are: the spinor field should be in the positive energy plane of the
boundary operator, that is, in the gauge transform of the positive energy
plane for the operator determined by $g=1.$ We assume that at the `initial
point' $g=1$ there are no zero modes. It follows that the operator $D_B$
does not have zero modes on the boundary $t=1.$ (Otherwise we could modify
$D(S^2)$ by adding a small positive constant.) The boundary conditions at
$t=0$ are the usual ones, i.e., the spinor field should be in the negative
energy plane of the 'free' Dirac hamiltonian.

The index formula (2.6) on manifolds with boundary contains two pieces.
The first is an integral of a local differential form in the interior of
the manifold. The $\eta$-invariant term is a nonlocal expression involving
the boundary Dirac operator. Because it  is expressed in terms of the
eigenvalues of the (hermitean) Dirac operator  it is invariant under gauge
transformations.

For this reason, when computing the index for the family of operators
given by the different gauge configurations, the only part contributing is
the local part. If $M$ is a sphere the relevant characteristic class is
the Chern class $c_{n+2}$ on $B.$ The Chern class $c_k$ is the coefficient
of $\l^k$ in the expansion of $\text{det}(1+\frac{\l}{2\pi i} F),$ where
$F$ is the curvature form. In the case of $G=SU(N)$, $\TR F=0$ and the
lowest terms are $$c_2= \frac{1}{8\pi^2}\TR F^2, c_3= \frac{i}{24\pi^3}\TR
F^3, c_4= \frac{1}{2^6\pi^4}( \TR F^4 - \frac12 (\TR F^2)^2.$$ The Chern
classes $c_n$ are normalized such that their integrals over closed
submanifolds of the corresponding dimension are integers.

The vector potential is globally defined and therefore the integral of the
Chern classes is given by a boundary integral of a Chern-Simons form
$CS_i(A)$ in $i=2n+3$ dimensions, $d(CS_i) = c_{n+2}.$ At the boundary
component $t=0$ the form vanishes. So the only contribution is
$$\int_{S^2\times M} CS_i(A(1,x,z)).\tag2.9$$ Performing only the $M$
integration gives a closed 2-form on $S^2.$ For example, when dim$M=1$ the
CS form is $\frac{1}{8\pi^2}\text{tr} (AdA+\frac23 A^3),$ and we get
$$\omega_A(X,Y) = \frac{1}{4\pi} \int_{S^1} \text{tr}\, A_{\phi} [X,Y],$$
the curvature at the point $A$ in the directions of infinitesimal gauge
transformations $X,Y.$ (Note the normalization factor $2\pi$ relating the
Chern class to the curvature formula.) This is not quite the central term
of an affine Kac-Moody algebra, but it is equivalent to it (in the
cohomology with coefficients in Map$(\Cal A,\Bbb C)).$ In other words,
there is a 1-form $\theta$ along gauge orbits in $\Cal A$ such that
$d\theta= \o-c ,$ where $$c(X,Y)= \frac{i}{2\pi} \int \TR X
\partial_{\phi} Y$$ is the central term of the Kac-Moody algebra,
considered as a closed constant coefficient 2-form on the gauge orbits.
There is a simple explicit expression for $\theta,$ $$\theta_A (X)=
\frac{i}{4\pi} \int \TR A X.$$

When dim$M=3$ the curvature (or equivalently, the Schwinger term) is
obtained from the five dimensional Chern-Simons form $$CS_5(A)=
\frac{i}{24\pi^3}\TR (A (dA)^2  + \frac32 A^3 dA + \frac35 A^5).$$ By the
same procedure as in the one dimensional case we obtain $$\omega_A(X,Y)=
\frac{i}{4\pi^2}\int \TR \left((AdA+dA\,A+A^3)[X,Y]+XdA\,YA -YdA\,XA
\right)  . $$ This differs from the FM cocycle, [FaSh], [Mi],
$$\omega'_A(X,Y)=\frac{i}{24\pi^2} \int \TR A (dX dY - dY dX)$$ by the
coboundary of $$\frac{-i}{24\pi^2}\int \TR (AdA +dA\,A+A^3) X .$$

The use of index theory for describing hamiltonian anomalies was suggested
by Nelson and Alvarez-Gaume in [NeAl]. However, in that paper the
appearance of Schwinger terms was not made clear.

\vskip 0.3in
\noindent{3. TWO TECHNICAL POINTS.}
\vskip 0.3in

The eventual aim of the discussion in Section 4 is to obtain
formulae for the Dixmier-Douady class in terms of de Rham forms
on subsets of the space of connections.
We need to first deal with two technical issues.

The first of these  shows that working with
the natural open cover of $\Cal A$ defined in section 2 is
equivalent to the bundle gerbe in [CaMu].

Define the disjoint union
 $$
 Y = \coprod U_\lambda \subset \Cal A \times \RM
 $$
 as the set of all $(A, \lambda) $ such that
$A \in U_\lambda$. We topologize $Y$ by giving
$\RM$ the discrete topology. Notice that as a set
$Y$ is just $\Cal A_0$ but the topology is different.
The identity map $Y \to \Cal A_0$ is continuous.
The projection  $Y \to \Cal A$ is a submersion.

In general if  we have a gerbe $Q \to Z^{[2]}$ where
$Z \to X$ is a  submersion and another submersion $ Y \to X $
and a fibre map $ f \colon Y \to Z$ we get an induced map
$f^{[2]} \colon Y ^{[2]} \to Z^{[2]}$ and the line bundle $Q$
pulls back to define a gerbe $(f^{[2]})^* ( Q)$ on $X $.
The gerbes $Q$ and $(f^{[2]})^* ( Q)$ are
{\it stably isomorphic}.  Generally we say two gerbes $G_1$ and
$G_2$ are stably isomorphic if there are trivial gerbes $T$ and $S$ such
that $G_1 \otimes S$ is isomorphic to $G_2 \otimes T$.
The notion of stable isomorphism wasn't understood at the time
of writing \cite{Mu} but  the definition of
the tensor product, trivial gerbe  and isomorphism
of gerbes are given there.
The point of defining stable isomorphism in this way is that
two gerbes are stably isomorphic if and only if they
have the same Dixmier-Douady class. Note that this
definition of stable isomorphism
is the same idea used in K-theory to define stable
isomorphism of vector bundles, i.e. we say two vector bundles
$E$ and $F$ are stably isomorphic if one
 can find two
trivial bundles $\RM^n$ and $\RM^m$ such that
$E \oplus \RM^n $ and $F \oplus \RM^m$ are isomorphic.

So returning to the case of interest the gerbes over $\Cal A_0$
with either topology on $\Cal A_0$ are
equivalent so we can work with either picture.
 An advantage of the open
cover picture is that the map $\delta$ introduced in
\cite{Mu} is then just the coboundary map in the
C\'ech de-Rham double complex. In the next section
$\Cal A_0$ can be interpreted in either sense.

For technical reasons explained below it is worth noting that we may work
with a denumerable cover from the very beginning. If we restrict $\l$ to
be rational then the sets $U_{\l}$ form a denumerable cover. It follows
that the intersections  $U_{\l\l'}= U_{\l}\cap U_{\l'}$ also form a
denumerable open cover. Similarly, we have an open cover by sets
$V_{\l\l'}=\pi(U_{\l\l'})$ on the quotient $X=\Cal A/\Cal G_e,$ where
$\Cal G_e $  is the group of \it based \rm gauge transformations $g$,
$g(p)=e=$ the identity at some fixed base point $p\in M.$  Here $\pi:\Cal
A\to X$ is the canonical projection.

The second technical point is the question of existence of partitions of
unity. This is one of the major  technical difficulties with working with
manifolds modelled on infinite dimensional vector spaces which are not
Hilbert spaces. We digress here to indicate how this problem is solved for
the case we are presently interested in. The main result in this theory
appears to be the theorem of \cite{Mil}.

\medskip
\noindent{\bf Theorem}.{ \it If $M$ is a Lindel\"of,
regular manifold modelled
on a topological vector space with enough smooth functions then
any open cover of $M$ has a refinement which admits a partition
of unity. }
\medskip

Before trying to prove this let us give some definitions. Lindel\"of means
any open cover has a countable subcover. Regular means any closed set and
a point not in it can be separated by disjoint open sets.  A topological
vector space $V$ has enough smooth functions if the collection of sets of
the form $U_f = \{ x \in V \mid f(x) > 0 \}$ where $f$ runs over all
smooth functions is a basis for the topology of $V$.  Another way
of saying this is that for every point $x \in V$ and open
set $U$ containing $x$ there is a smooth function $f$
with $ x \in U_f \subset U$.

The reason to worry about not having enough smooth functions is that
the obvious method of constructing them,
 by taking a semi-norm $\rho$ and composing
it with a bump function on $\RM$, may not work as the   semi-norm may not
be smooth. However \cite{Mil, Be} show that
 the set of  smooth functions from a manifold
into a Hilbert space with the smooth, Fr\'echet topology has enough smooth
functions.  The point is that we can realize this topology by semi-norms
$\rho_k$
which are inner products defined by summing the $L^2$ norms
of the first $k$ derivatives. Then each of these is smooth because the
inner products are bilinear and hence smooth. That this gives rise to
the same topology as the uniform norms on derivatives is a consequence
of the Sobolev inequalities.

The proof of the theorem above from \cite{Mil} goes as follows.
Let $U$ be an element of a given open cover and let $x\in U.$
By the assumption, there is a smooth real valued function $f$ on $M$ such
that $x\in U_f \subset U.$ We can assume that $f$ is nonnegative. Namely,
for a given $f$ we can form another smooth function $\tilde f=h\circ f,$
where $h:\Bbb R\to \Bbb R$ is the smooth function defined by $h(x)=0$ for
$x\leq 0$ and $h(x)=\exp(-1/x^2)$ for $x>0.$ Clearly $U_{\tilde f}=U_f$
and $\tilde f$ is nonnegative. Choosing nonnegative functions has the
advantage that
$$
U_f \cap U_g = U_{fg}.        \tag3.1
$$

By the above discussion we can
refine the given open cover to an open cover consisting of $U_f$'s.
Then we can take
a countable open subcover $U_i = U_{f_i}$ for some smooth, non-negative
functions $f_i$.
Consider now the sets
$$
V_n = \{ x \mid f_1(x) < 1/n\} \cap \{ x \mid f_2(x) < 1/n\} \cap \cdots
   \{ x \mid f_{n-1}(x) < 1/n\} \cap U_n.
$$
These are a locally finite cover.

First let us prove they are a cover.
Note that the  $U_n$ cover $M$ so
there is some $f_m$ such that $f_m(x) > 0$.
Let $x \in M$ and assume that $f_k$ is the first function
which doesn't vanish at $x$. Then $x \in V_k$.

To see that  this cover is  locally finite  pick a point $x \in M$ and some
$f_m$ such that $f_m(x) > 0$. But then the only possible $V_n$ that can
contain $x$ are those where $ n < 1/f_m(x)$. Clearly we can find an open
set around $x$ where $f_m(x)$ stays positive and bounded so a similar
result holds for all the points in that open set so the cover $V_n$ has to
be locally finite. Each of the sets $\{x| f_k(x) < i/n \}$ can be written
as $U_f$ for some suitable nonnegative function $f$ (essentially the same
argument as before (3.1)). {}From (3.1) it follows that each $V_n$ is of the
form $U_{g_n}$ for some smooth functions $g_n$. The partitions of unity
are obtained by scaling the $g_n$ by their sum.

In the case at hand  we can see directly that the open cover we are using
is of the form required by the preceding construction of the partition of
unity. This is because we can define smooth functions $f_{\l\l'}$ on $X$
as $f_{\l\l'}(A)=\exp(-1/d),$ where $d$ is the distance of the spectrum of
the operator $D_A$ to the set $\{\l,\l'\}.$ This distance is always
positive for $A\in U_{\l\l'},$ because the spectrum does not have
accumulation points on a compact manifold $M.$ When $A\notin U_{\l\l'}$ we
set $f_{\l\l'}(A)=0.$ Finally $\Cal A$ and $\Cal A/\Cal G_e$ are metric
spaces as they are Frechet manifolds modelled on a space with topology
given by the (countably many) Sobolev space inner products and hence are
regular. We can use the set of functions $f_{\l\l'}$ in the proof above to
show the existence of a locally finite cover and corresponding partition
of unity.

\vskip 0.3 in
\noindent {4. CALCULATING THE DIXMIER-DOUADY CLASS.}
\vskip 0.3 in

Our starting point is the gerbe $J$ over $\Cal A$ defined in [CaMu].
This is a line bundle over the fibre product ${\Cal A}_0^{[2]}$. This
fibre product can be identified with all triples $(A, \l, \l')$
where neither $\l$ nor $\l'$ are in the spectrum of $D_A$. The fibre
of $J$ over $(A, \l, \l')$ is just $DET_{\l\l'}$.
Let $\pi \colon {\Cal A}_0 \to \Cal A$
be the projection. Let $p \colon \Cal A \to \Cal A/\Cal G_e$ be the quotient
by the gauge action.
We saw in \cite{CaMu1} that the line bundle $DET$ on $\Cal A_0$ satisfies
$J = \delta(DET)$. Here $\delta(DET) = \pi_1^*(DET)^*\otimes \pi_2^*(DET)$
where $\pi_i \colon \Cal A_0^{[2]} \to \Cal A_0$ are the projections,
$$
\pi_1((A, \l, \l')) = (A,  \l) \quad\text{and}\quad
\pi_2((A, \l, \l')) = (A,  \l')
$$
In otherwords $J=\delta(DET)$ is equivalent to
$$
DET_{\l\l'} = DET_{\l}^*\otimes DET_{\l'}
$$
which is equivalent to equation $(2.2)$. Note that we also used $\delta$ to
denote a similar operation on differential forms discussed earlier.

The fibering ${\Cal A}_0 \to \Cal A$ has, over each open set $U_\lambda$
a canonical section $A \mapsto (A, \lambda)$.
These enable us to suppress the geometry of the fibration
and the gerbe $J$ becomes the line bundle $DET_{\lambda\l'}$
over the intersection $U_{\l\l'}$ and its triviality amounts
to the fact that we have the line bundle $DET_\l $ over $U_\l$
and over intersections we have the identifications
$$
DET_{\l\l'} = DET_{\l}^*\otimes DET_{\l'}.
$$

We denote the Chern class of $DET_{\l\l'}$ by $\theta_2^{[2]}$.
Note that these bundles descend to
bundles over $V_{\l\l'}=\pi(U_{\l\l'})\subset \Cal A/\Cal G_e.$
Therefore, the forms $\theta_2^{\l\l'}=\theta_2^{\l}-\theta_2^{\l'}$
on $U_{\l\l'}$ (where $\theta_2^{\l}$ is the 2-form giving the
curvature of $DET_{\l}$) are equivalent (in cohomology) to forms
which descend to closed 2-forms $\phi_2^{\l\l'}$ on $V_{\l\l'}.$

Our aim in this section is twofold. We show first that the collection of
Chern classes $\theta_2^{\lambda}$ gives rise to the Dixmier-Douady class
of the bundle gerbe $J/\Cal G_e$ and second that using the results of the
preceding section, we can obtain formulae for this class using standard
methods.

To begin, let us choose a bundle gerbe connection on $J/\Cal G$. This is
possible by orthogonal projection. Call it $\nabla$ and its curvature
$F_\nabla$. Then we can pull $\nabla $ back to $p^*(\nabla)$ on $J$ with
curvature $p^*(F_\nabla)$.  Similarly choose a connection $\nabla_{DET}$
on the line bundle $DET$. This induces a connection $\delta(\nabla_{DET})$
on $J$.  The difference of these two connections is a one form $a$ on
$\Cal A_0^{[2]}$ and, in fact, $\delta(a) = 0$ so that $a = \delta(\psi)$.

Note that $\psi$ is not unique and we do not have a constructive method of
finding it (but if we did then we could construct explicit formulae).
Pressing on however if $F_{DET}$ is the curvature (which has class equal
to the  chern class $\{\theta_2^\lambda\}$) of $\nabla_{DET}$ then
$$
p^*(F_\nabla) = \delta(F_{DET} + d\psi).
$$

Now assume that down on $\Cal A/\Cal G_e$ we have solved
$$
F_{\nabla} = \delta(f).
$$

We remark that  this is a central point. It is not obvious that there is a
solution. However by the previous subsection there is a locally finite
partition of unity $\{s_{\lambda}\}$ subordinate to the open cover
$\{V_{\lambda}\}.$ The curvature of the gerbe consists of closed 2-forms
$\phi_2^{\lambda\lambda'}$ on the intersections $V_{\lambda\lambda'}$
satisfying the cocycle condition
$$
\phi_2^{\lambda\lambda'} +\phi_2^{\lambda'\lambda''}=
\phi_2^{\lambda\lambda''
}
$$
on the domains of definition. One can then define
$$\phi_2^{\lambda} = \sum s_{\lambda'} \phi_2^{\lambda\lambda'}$$
which gives
$$\phi_2^{\lambda\lambda'}= \phi_2^{\lambda}-\phi_2^{\lambda'}$$
on $V_{\lambda\lambda'}.$
The collection of forms $\phi_2^{\lambda}$ defines the
form $f$ on $\Cal A_0/\Cal G_e.$

Now, continuing our argument,
 we have
$$
\delta(p^*(f) ) = \delta(F_{DET} + d\psi)
$$
so that
$$
p^*(f)  = F_{DET} + d\psi  + \pi^*(\rho)
$$
as $\pi^*(\rho) = \delta(\rho)$.

By definition
the Dixmier-Douady class is the 3-form $\omega$ on $\Cal A/\Cal G_e$
defined by
$$d f = \pi^*(\omega).\tag4.1$$
If we transgress $\omega$ then we want to solve
$$
p^*(\omega) = d\mu
$$
for some $\mu$. Hence we have
$$
p^*(\pi^*(\omega)) = d\pi^*(\mu).
$$
But from the above we have
$$
p^*\pi^*(\omega) = d p^*(f)  = dF_{DET} + dd\psi +  d\pi^*(\rho)  =
\pi^*(d\rho).
$$
Hence
$$
p^*(\omega) = d \rho.
$$

So $[\rho]=[\mu]$. To calculate the Lie algebra cocycle we need to apply
$\rho$ to two vectors $\xi$, $\eta$ in $\Cal A$ generated by the group
action. As the group also acts on $\Cal A_0$ it is equivalent to apply
$\pi^*(\rho)$ to two such vectors which we shall denote by the same
symbols. Then, noting that $p^*(f)$ is  zero on any vectors generated by
the gauge group action (because $p$ is the projection $\Cal A \to \Cal
A/\Cal G_e$)  we have
$$
\pi^*(\rho)(\xi, \eta)  = - F_{DET}(\xi, \eta)  - d\psi(\xi, \eta).
$$

Hence the  Faddeev-Mickelsson cocycle on the Lie algebra of the gauge group
is cohomologous to the negative of that defined by the curvature $F_{DET}$
of the line bundle $DET$.

To obtain the Dixmier-Douady class as a characteristic class
we recall that
in the case of even dimensional manifolds, Atiyah and Singer \cite{AtSi}
gave a construction of `anomalies' in terms of characteristic classes.
In the present case of odd dimensional manifolds we
now demonstrate that a similar procedure yields the Dixmier-Douady class.

We begin with the observation that
given a closed integral form $\Omega_p$ of degree $p$ on a product manifold
$M\times X$ (dim$M=d$ and dim$S=k$) we obtain a closed integral form on
$S,$ of degree $p-d,$ as
$$\Omega_X = \int_M \Omega.$$
If now $A$ is any  Lie algebra valued connection on the product $M\times X$
and $F$ is the corresponding curvature we can construct the Chern form
$c_{2n} = c_{2n}(F)$ as a polynomial in $F.$ Apply this to the connection
$A$ defined by Atiyah and Singer, [AtSi], [DoKr p. 196], in the case when
$X=\Cal A/\Cal G_e.$

First pull back the forms to $M\times \Cal A.$ The Atiyah-Singer
connection on $M\times X$ becomes a globally defined Lie algebra valued
1-form $\hat A$ on $M\times \Cal A.$ Along directions $u$ on $M$ it is
defined as $$\hat A_{x,a}(u)= u_{\mu} a_{\mu}(x), \text{ with $x\in M$ and
$a\in\Cal A$}$$ and along a tangent vector $b\in T_a \Cal A$ the value is
$$\hat A_{x,a}(b) = -(G_a d_a^* b)(x) ,$$ where $d_a = d +[a,\cdot]$ is
the covariant exterior differentiation acting on functions with values in
the adjoint representation of $\gm$ and $G_a= (d_a^* d_a)^{-1}$ is the
Green's operator. Let $\hat F$ be the curvature form determined by $\hat
A.$ A tangent vector $b$ at $a\in\Cal A$ is said to be in the \it
background gauge \rm if $d_a^* b= \partial_{\mu} b_{\mu} +
[a_{\mu},b_{\mu}] =0.$  Any tangent vector $b$ at a point $\pi(a)\in X$ is
represented by a unique potential $b$ in the background gauge. For this
reason we need to evaluate the curvature $\hat F$ only along background
gauge directions.

Along tangent vectors $u,v$ at $x\in M$ the curvature is $\Cal F_{
x,a}(u,v) = f_x(u,v),$ where $f= da +\frac12 [a,a].$ Along directions
$b,b'$ in the  background gauge at  $a\in \Cal A$ the value of $\hat F$ is
$G_a[b_{\mu}, b_{\mu}']$ and finally along mixed directions $\hat
F_{x,a}(u,b)= u_{\mu} b_{\mu}(x),$ [AtSi]. For example, when dim$M=3$ and
$p=6$ the 3-form $\Omega_X$ becomes now, evaluated at $a\in\Cal A,$
$$\align \Omega_X(b,b',b'') &=  \frac{-i}{8\pi^3}\int_M \TR b[b',b''] +\\
&\frac{i}{8\pi^3}\int_M \TR f (b'' G_a[b,b'] + G_a[b,b'] b''+ \text{ cycl.
combin. }),\endalign $$ when $b,b',b''$ are in the background gauge.

The integral of $\Omega_X$ over a sphere $S^3\subset X$ can be evaluated
without computing the nonlocal Green's operators in the above formula. The
pull-back of  $S^3$  becomes a disk $D^3$ on $\Cal A$ with boundary points
identified through gauge transformations. We can therefore write
$$\int_{S^3} \Omega_X = \int_{M\times D^3} c_{2n}(\hat F).$$ But the
integral of the Chern form over a manifold with a boundary (when the
potential is globally defined) is equal to the integral $$\int_{M\times
\partial D^3} CS_{2n-1}(\hat A).$$ Along gauge directions the form $\hat
A$ is particularly simple: $\hat A_{a,x} (b)=Z(x),$ where $x\in M, a\in
\Cal A,$ and $ b=-d_a Z=[Z,a] - dZ$ is a tangent vector along a gauge
orbit at $a.$ For example, when $M=S^1$ and $2n=4$ we get  (here
$S^2=\partial D^3$) $$\int_{S^3} \Omega_X = \int CS_3(\hat A) =
\frac{1}{8\pi^2}\int_{S^1\times S^2} \text{tr} (a^g d a^g +\frac23
(a^g)^3) = \frac{1}{24\pi^2}\int_{S^1\times S^2} \text{tr} (dg g^{-1})^3$$
where $g=g(x,z)$ is a family of gauge transformations parametrized by
$z\in S^2.$  Similar results hold in higher dimensions: The exponent 3 on
the right is replaced by dim$M+2=2n+3$ and the normalization factor is
$-(\frac{i}{2\pi})^{n+2} ((n+2)!\cdot (2n+3))^{-1}.$

Now we can prove that $\Omega_X$ represents the Dixmier-Douady class of
the bundle gerbe. The integral of the DD form $\omega$ over a closed
3-cycle $S\subset \Cal A/\Cal G_e$ (which can be assumed to be a sphere
$S^3$) is evaluated, using the pull-back form $df=\pi^*(\omega),$ on $\Cal
A_0/\Cal G_0$ and further pulling back this by $p^*$ to $\Cal A_0$ But the
cover of $S^3$ in the latter space is a disk $D^3$ such that the boundary
points are gauge related. Because the spectrum of the Dirac operator is
gauge invariant we can choose a single label $\l$ such that $\partial D^3
\subset U_{\l}.$ Since $p^*(df)= d(p^*f)$ the integral over $D^3$ can be
evaluated by Stokes theorem over the boundary $\partial D^3.$ The form
$p^*f$ on $U_{\l}$ is equal to ${\theta_2}^{\l},$ the Chern class of the
determinant bundle over $U_{\l}.$ But the integral of ${\theta_2}^{\l}$
over the gauge orbit $\partial D^3$ is given by the integral of the
Chern-Simons form (2.9), thus giving exactly the same result as the
integration of $\Omega_X$ above. We conclude that the de Rham cohomology
classes $[\omega]$ and $[\Omega_X]$ are the same.

\vskip 0.3in
\noindent{REFERENCES.}

\vskip 0.3in

\noindent  [AtSi]  M.F. Atiyah and I.M. Singer: Dirac operators coupled to
vector
potentials. Natl. Acad. Sci. (USA) \bf 81, \rm 2597 (1984)

\noindent[AtPaSi] M.F. Atiyah, V.K. Patodi, and I. M. Singer: Spectral
asymmetry and
Riemannian Geometry, I-III. Math. Proc. Camb. Phil. Soc. \bf 77, \rm 43
(1975); \bf 78, \rm 405 (1975); \bf 79, \rm 71 (1976).

\noindent[Be] Edwin J. Beggs: {\it The de Rham complex on
infinite dimensional manifolds.}  Quart. J. Math. Oxford (2),
38 (1987), 131-154.

\noindent[Br] J.-L. Brylinski: \it Loop Spaces, Characteristic Classes and
Geometric Quantization. \rm Birkh\"auser, Boston-Basel-Berlin (1993)

\noindent[CaMu]  A.L. Carey and M.K. Murray.:
Mathematical remarks on the cohomology of gauge groups
and anomalies. To appear in Int J. Mod. Phys. A. hep-th/9408141.

\noindent[CaMu1]  A.L. Carey and M.K. Murray.:
Faddeev's anomaly and bundle gerbes.
To appear in Lett. Math. Phys. 1995.

\noindent[CaMuWa] A.L. Carey, M.K. Murray and B. Wang.:
Higher bundle gerbes, descent equations and 3-Cocycles
preprint 1995.

\noindent[DiDo] J. Dixmier and A. Douady: Champs continus d'espaces
hilbertiens et de
$C^*$ algebres. Bull. Soc. Math. Fr. \bf 91, \rm 227 (1963)

\noindent[DoKr] S.K. Donaldson and P.B. Kronheimer: \it The Geometry of
Four-Manifolds.
\rm Clarendon Press, Oxford (1990)

\noindent[FaSh] L. Faddeev and S. Shatasvili: Algebraic and hamiltonian
methods in the
theory of nonabelian anomalies. Theoret. Math. Phys. \bf 60, \rm
770 (1985)

\noindent[Mi] J. Mickelsson: Chiral anomalies in even and odd dimensions.
Commun.
Math. Phys. \bf 97, \rm 361 (1985)

\noindent[Mi1] J. Mickelsson: On the hamiltonian approach to commutator
anomalies in $3+1$ dimensions. Phys. Lett. \bf B 241, \rm 70 (1990)

\noindent[Mil]  J. Milnor: {\it On infinite dimensional Lie groups}.
Preprint.

\noindent[Mu] M.K. Murray.: {\sl Bundle gerbes.}
dg-ga/9407015. To appear in  Journal of  the London Mathematical Society.

\noindent[NeAl] P. Nelson and L. Alvarez-Gaume: Hamiltonian interpretation
of anomalies.
Commun. Math. Phys. \bf 99, \rm 103 (1985)

\noindent[PrSe] A. Pressley and G. Segal: \it Loop Groups. \rm Clarendon
Press, Oxford
(1986)

\noindent[Se] G. Segal:  Faddeev's anomaly
in Gauss's law. Preprint.

\enddocument